\documentclass[sigconf]{acmart}
\renewcommand\footnotetextcopyrightpermission[1]{} 
\usepackage{booktabs} 
\usepackage{amsmath}
\usepackage{amssymb}
\usepackage{algpseudocode,algorithm,algorithmicx}
\usepackage[capitalise]{cleveref}
\usepackage{varwidth}
\usepackage{siunitx}

\newcommand*{\symDefine}[2]{\newcommand{{#1}}{{#2}}}
\newcommand\symVariance{\operatorname{Var}}
\newcommand\symExpectation{\operatorname{E}}
\newcommand\symCovariance{\operatorname{Cov}}

\symDefine{\symInputIndex}{i}
\symDefine{\symInputSize}{n}
\symDefine{\symInputElement}{d}
\symDefine{\symInputElementMin}{\symInputElement'}
\symDefine{\symInputSet}{D}
\symDefine{\symHashSize}{m}
\symDefine{\symIntermediateHash}{r}
\symDefine{\symHashElement}{h}
\symDefine{\symHashIndex}{j}
\symDefine{\symHashIndexTwo}{k}
\symDefine{\symJaccard}{J}
\symDefine{\symJaccardEstimator}{\hat{J}}
\symDefine{\symProbability}{P}
\symDefine{\symProbabilityDensity}{\rho}
\symDefine{\symShuffleArrayStamp}{q}
\symDefine{\symShuffleArrayPos}{p}
\symDefine{\symProbabilityMass}{\rho}
\symDefine{\symShuffleIndex}{k}
\symDefine{\symHistogram}{b}
\symDefine{\symMaxNonZero}{a}
\symDefine{\symMaxNonZeroValue}{s}
\symDefine{\symSetA}{A}
\symDefine{\symSetB}{B}
\symDefine{\symSetS}{S}
\symDefine{\symNewPosition}{p}
\symDefine{\symPermutation}{\pi}
\symDefine{\symBigO}{\mathcal{O}}
\symDefine{\symRuntime}{t}
\symDefine{\symRunningTime}{T}
\symDefine{\symIndicator}{I}
\symDefine{\symHarmonicNumber}{H}
\symDefine{\symImprovement}{\alpha}
\symDefine{\symUnionSize}{u}
\symDefine{\symSumIndex}{l}

\DeclareMathOperator*{\argmin}{arg\,min}

\setcopyright{none}

\begin{document}
\sloppy
\allowdisplaybreaks
\title{SuperMinHash -- A New Minwise Hashing Algorithm\\for Jaccard Similarity Estimation}

\author{Otmar Ertl}
\affiliation{%
  \city{Linz} 
  \country{Austria} 
}
\email{otmar.ertl@gmail.com}

\begin{abstract}
This paper presents a new algorithm for calculating hash signatures of sets which can be directly used for Jaccard similarity estimation. The new approach is an improvement over the MinHash algorithm, because it has a better runtime behavior and the resulting signatures allow a more precise estimation of the Jaccard index.
\end{abstract}

\maketitle

\section{Introduction}
The Jaccard index
\begin{equation*}
\symJaccard
=
\frac{\left| \symSetA \cap\symSetB \right|}{\left| \symSetA \cup\symSetB \right|}
\end{equation*}
is a measure for the similarity of two sets $\symSetA$ and $\symSetB$. If one is interested in pairwise similarities of many sets the direct calculation is often computationally too expensive. Therefore, different algorithms \cite{Broder1997,Li2012, Shrivastava2014, Shrivastava2014a, Shrivastava2017, Dahlgaard2017} have been proposed, which first calculate hash signatures of individual sets. The Jaccard index can then be quickly determined given only the signatures of the corresponding two sets. Each signature contains condensed information about its corresponding set which is relevant for Jaccard index estimation.

\subsection{MinHash Algorithm}

\begin{algorithm}[b]
\caption{MinHash algorithm.}
\label{alg:minwise}
\begin{algorithmic}
\Require $(\symInputElement_0,\symInputElement_1,\ldots,\symInputElement_{\symInputSize-1})$
\Ensure $(
\symHashElement_0,
\symHashElement_1,
\ldots,
\symHashElement_{\symHashSize-1}
)\in [0, 1)^\symHashSize$
\State $(
\symHashElement_0,
\symHashElement_1,
\ldots,
\symHashElement_{\symHashSize-1}
)\gets
(\infty,\infty,\ldots,\infty)$

\For{$\symInputIndex \gets 0,1,\ldots, \symInputSize-1$}
\State initialize pseudo-random generator with seed $\symInputElement_\symInputIndex$
\For{$\symHashIndex \gets 0,1,\ldots, \symHashSize-1$}
\State $\symIntermediateHash\gets$ uniform random number from $[0,1)$
\State $\symHashElement_\symHashIndex \gets \min(\symHashElement_\symHashIndex, \symIntermediateHash)$
\EndFor
\EndFor
\end{algorithmic}
\end{algorithm}

The MinHash algorithm \cite{Broder1997} was the first approach to calculate signatures suitable for Jaccard index estimation. The signature consists of $\symHashSize$ values $(\symHashElement_0,\symHashElement_1,\ldots,\symHashElement_{\symHashSize-1})$ which are defined for a given data set $\symInputSet$ by
\begin{equation}
\label{equ:def_minwise_hashing}
\symHashElement_\symHashIndex(\symInputSet)
:=
\min_{\symInputElement\in\symInputSet}(
\symIntermediateHash_{\symHashIndex}(\symInputElement)
).
\end{equation}
The functions $\symIntermediateHash_{\symHashIndex}$ are independent and uniform hash functions with value range $[0,1)$.  The signature size $\symHashSize$ is a free parameter and allows trading space and computation time for more precise estimates.

The probability that signature values are equal for two different sets $\symSetA$ and  $\symSetB$ corresponds to the Jaccard index
\begin{equation}
\label{equ:prop_equ_sketch}
\symProbability(\symHashElement_\symHashIndex(\symSetA) = \symHashElement_\symHashIndex(\symSetB))
=
\symProbability(\symHashElement_\symHashIndex(\symSetA\cap\symSetB)
=
\symHashElement_\symHashIndex(\symSetA\cup\symSetB))
=
\frac{\left| \symSetA \cap\symSetB \right|}{\left| \symSetA \cup\symSetB \right|}
=
J.
\end{equation}
Here we used the equivalence 
$\symHashElement_\symHashIndex(\symSetA) = \symHashElement_\symHashIndex(\symSetB) \Leftrightarrow \symHashElement_\symHashIndex(\symSetA\cap\symSetB)
=
\symHashElement_\symHashIndex(\symSetA\cup\symSetB)$.
Therefore, 
\begin{equation}
\label{equ:jaccard_estimator}
\symJaccardEstimator = \frac{1}{\symHashSize} \sum_{\symHashIndex=0}^{\symHashSize-1}\symIndicator(\symHashElement_\symHashIndex(\symSetA) = \symHashElement_\symHashIndex(\symSetB))
\end{equation}
is an unbiased estimator for the Jaccard index. $\symIndicator$ denotes the indicator function. Since all signature values are independent and identically distributed, the sum of indicators corresponds to a binomial distribution with sample size $\symHashSize$ and success probability $\symJaccard$. Hence, the variance of the estimator is given by
\begin{equation}
\label{equ:minwise_hashing_variance}
\symVariance(\symJaccardEstimator)
=
\frac{\symJaccard(1-\symJaccard)}{\symHashSize}.
\end{equation}

\cref{alg:minwise} demonstrates the calculation of the MinHash signature for a given input data sequence $\symInputElement_0, \symInputElement_1,\ldots,\symInputElement_{\symInputSize-1}$ of length $\symInputSize$. Since the input data may contain duplicates we generally have $|\symInputSet|\leq \symInputSize$ for the cardinality of the set $\symInputSet = \lbrace \symInputElement_0, \symInputElement_1,\ldots,\symInputElement_{\symInputSize-1}\rbrace$.

For simplicity \cref{alg:minwise} and also the algorithms that are presented later are expressed in terms of a pseudo-random number generator. Assuming independent and uniform hash functions $\symIntermediateHash_\symHashIndex$ the sequence $\symIntermediateHash_{0}(\symInputElement), \symIntermediateHash_{1}(\symInputElement),\ldots$ behaves statistically like the output of an ideal pseudo-random generator with seed $\symInputElement$. By chaining the hash values of different hash functions random bit sequences of arbitrary length can be realized. In practice, the next hash function is evaluated, only if all bits of the previous hash value have been consumed.

The runtime complexity of MinHash is $\symBigO(\symHashSize \symInputSize)$, because the inner loop is executed $\symHashSize \symInputSize$ times. Since $\symHashSize$ is large for many applications, more efficient algorithms are desirable.

\subsection{One Permutation Hashing}
The first approach that significantly reduced the calculation time was one permutation hashing \cite{Li2012}. The idea is to divide the input set $\symInputSet$ randomly into $\symHashSize$ disjoint subsets $\symInputSet_0, \symInputSet_1, \ldots, \symInputSet_{\symHashSize-1}$. The hash signature is calculated using a single hash function $\symIntermediateHash$
\begin{equation*}
\symHashElement_\symHashIndex(\symInputSet)
:=
\min_{\symInputElement\in\symInputSet_\symHashIndex}(
\symIntermediateHash(\symInputElement)
).
\end{equation*}
This procedure results in an optimal runtime complexity of $\symBigO(\symHashSize + \symInputSize)$. Unfortunately, for small input sets, especially if $|\symInputSet| < \symHashSize$, many subsets are empty and corresponding signature values are undefined. Various densification algorithms have been proposed to resolve this problem  \cite{Shrivastava2014, Shrivastava2014a, Shrivastava2017}, which fill undefined positions in the signature by copying defined values in such a way that estimator \eqref{equ:jaccard_estimator} remains unbiased. However, 
all densified hash signatures lead to less precise Jaccard index estimates compared to MinHash for small data sets with $|\symInputSet|\ll \symHashSize$. In addition, the best  densification scheme in terms of precision presented in \cite{Shrivastava2017} has a runtime that scales quadratically with signature size $\symHashSize$ for very small data sets \cite{Dahlgaard2017}. Another disadvantage is that signatures of different sets cannot be longer merged after densification to construct the signature for the corresponding union set.

\subsection{Fast Similarity Sketching}
Recently, a new algorithm called fast similarity sketching has been presented \cite{Dahlgaard2017} that achieves a runtime complexity of $\symBigO(\symInputSize + \symHashSize \log\symHashSize)$ for the case that the input does not contain duplicates ($\symInputSize = |\symInputSet|$). It was also shown that the variance of the Jaccard index estimator is significantly improved for small data sets. However, in contrast to MinHash it cannot be directly used as streaming algorithm, because multiple passes over the input data are needed. Moreover, the computation time is approximately twice that of MinHash for small data sets with $|\symInputSet| \ll \symHashSize$. 

\subsection{Outline}
In the following we present a new algorithm for the calculation of signatures appropriate for Jaccard index estimation. We call the new algorithm SuperMinHash, because it generally supersedes MinHash. We will prove that the variance of the Jaccard index estimator \eqref{equ:jaccard_estimator} is strictly smaller for same signature sizes. In addition, we will show that the runtime for calculating the signatures is comparable for small data sets while it is significantly better for larger data sets as it follows an $\symBigO(\symInputSize + \symHashSize \log^2 \symHashSize)$ scaling law for $\symInputSize = |\symInputSet|$. Furthermore, like MinHash, the new algorithm requires only a single pass over the input data, which allows a straightforward application to data streams or big data sets that do not fit into memory as a whole.

\section{SuperMinHash Algorithm}

\begin{algorithm}[b]
\caption{Straightforward calculation of the new signature defined by \eqref{equ:def_new_hash} using Fisher--Yates shuffling.}
\label{alg:plain_new}
\begin{algorithmic}
\Require $(\symInputElement_0,\symInputElement_1,\ldots,\symInputElement_{\symInputSize-1})$
\Ensure $(
\symHashElement_0,
\symHashElement_1,
\ldots,
\symHashElement_{\symHashSize-1}
)\in [0, \symHashSize)^\symHashSize$
\State $(
\symHashElement_0,
\symHashElement_1,
\ldots,
\symHashElement_{\symHashSize-1}
)\gets
(\infty,\infty,\ldots,\infty)$

\For{$\symInputIndex \gets 0,1,\ldots, \symInputSize-1$}
\State initialize pseudo-random generator with seed $\symInputElement_\symInputIndex$
\State $(\symShuffleArrayPos_0,\symShuffleArrayPos_1,\dots,\symShuffleArrayPos_{\symHashSize-1})\gets (0,1,\ldots,\symHashSize-1)$
\For{$\symHashIndex \gets 0,1,\ldots, \symHashSize-1$}
\State $\symShuffleIndex \gets$ uniform random  number from $\lbrace \symHashIndex, \ldots, \symHashSize-1\rbrace$
\State swap $\symShuffleArrayPos_\symHashIndex$ and $\symShuffleArrayPos_\symShuffleIndex$
\EndFor
\For{$\symHashIndex \gets 0,1,\ldots, \symHashSize-1$}
\State $\symIntermediateHash\gets$ uniform random number from $[0,1)$
\State $\symHashElement_\symHashIndex \gets \min(\symHashElement_\symHashIndex, \symIntermediateHash + \symShuffleArrayPos_\symHashIndex)$
\EndFor
\EndFor
\end{algorithmic}
\end{algorithm}

The new algorithm is based on a hash  signature defined by
\begin{equation}
\label{equ:def_new_hash}
\symHashElement_\symHashIndex(\symInputSet)
:=
\min_{\symInputElement\in\symInputSet}(
\symIntermediateHash_{\symHashIndex}(\symInputElement) + \symPermutation_{\symHashIndex}(\symInputElement)
).
\end{equation}
Here we extended \eqref{equ:def_minwise_hashing} by adding elements of a random permutation
\begin{equation*}
\symPermutation(\symInputElement) =
\begin{pmatrix}
0 & 1 & \cdots & \symHashSize-1 \\
\symPermutation_0(\symInputElement) & \symPermutation_1(\symInputElement) & \cdots &\symPermutation_{\symHashSize-1}(\symInputElement)
\end{pmatrix}
\end{equation*}
that is generated for each input element $\symInputElement$.

Since the values $
\symIntermediateHash_{\symHashIndex}(\symInputElement_0) + \symPermutation_{\symHashIndex}(\symInputElement_0),
\ldots,
\symIntermediateHash_{\symHashIndex}(\symInputElement_{\symInputSize-1}) + \symPermutation_{\symHashIndex}(\symInputElement_{\symInputSize-1}),
$ are still mutually independent and uniformly distributed over $[0,\symHashSize)$, \eqref{equ:prop_equ_sketch} also holds here and the Jaccard index estimator \eqref{equ:jaccard_estimator} will give unbiased results. However, in contrast to MinHash, the signature values $\symHashElement_0,\symHashElement_1,\ldots,\symHashElement_{\symHashSize-1}$ are no longer independent. As we will see, this is the reason for the improved precision when estimating the Jaccard index for small sets.

The new approach requires the generation of random permutations for each input data element. Fisher--Yates shuffling is the standard algorithm for this purpose \cite{Fisher1938}. The shuffling algorithm uses uniformly distributed integer numbers. An algorithm for the generation of strict uniform random integers that is efficient regarding random bit consumption is described in \cite{Lumbroso2013}.

A straightforward implementation of \eqref{equ:def_new_hash} would look like \cref{alg:plain_new}. Obviously, the runtime complexity is still $\symBigO(\symInputSize \symHashSize)$. However, in the following we describe a couple of algorithmic optimizations which finally end up in the new SuperMinHash algorithm.

\subsection{Optimization}

\begin{algorithm}[b]
\caption{Transformed version of \cref{alg:plain_new}.}
\label{alg:modified_new}
\begin{algorithmic}
\Require $(\symInputElement_0,\symInputElement_1,\ldots,\symInputElement_{\symInputSize-1})$
\Ensure $(
\symHashElement_0,
\symHashElement_1,
\ldots,
\symHashElement_{\symHashSize-1}
)\in [0, \symHashSize)^\symHashSize$
\State $(
\symHashElement_0,
\symHashElement_1,
\ldots,
\symHashElement_{\symHashSize-1}
)\gets
(\infty,\infty,\ldots,\infty)$
\State allocate array $(\symShuffleArrayPos_0,\symShuffleArrayPos_1,\dots,\symShuffleArrayPos_{\symHashSize-1})$
\State $(\symShuffleArrayStamp_0,\symShuffleArrayStamp_1,\dots,\symShuffleArrayStamp_{\symHashSize-1})\gets (-1,-1,\ldots,-1)$

\For{$\symInputIndex \gets 0,1,\ldots, \symInputSize-1$}
\State initialize pseudo-random generator with seed $\symInputElement_\symInputIndex$
\For{$\symHashIndex \gets 0,1,\ldots, \symHashSize-1$}
\State $\symIntermediateHash\gets$ uniform random number from $[0,1)$
\State $\symShuffleIndex \gets$ uniform random  number from $\lbrace \symHashIndex, \ldots, \symHashSize-1\rbrace$
\If{$\symShuffleArrayStamp_\symHashIndex\neq \symInputIndex$}
\State $\symShuffleArrayStamp_\symHashIndex\gets\symInputIndex$
\State $\symShuffleArrayPos_\symHashIndex\gets\symHashIndex$
\EndIf
\If{$\symShuffleArrayStamp_\symShuffleIndex\neq \symInputIndex$}
\State $\symShuffleArrayStamp_\symShuffleIndex\gets\symInputIndex$
\State $\symShuffleArrayPos_\symShuffleIndex\gets\symShuffleIndex$
\EndIf
\State swap $\symShuffleArrayPos_\symHashIndex$ and $\symShuffleArrayPos_\symShuffleIndex$
\State $\symHashElement_{\symShuffleArrayPos_\symHashIndex} \gets \min(\symHashElement_{\symShuffleArrayPos_\symHashIndex}, \symIntermediateHash + \symHashIndex)$
\EndFor
\EndFor
\end{algorithmic}
\end{algorithm}

\begin{algorithm}[b]
\caption{SuperMinHash algorithm which is an optimized version of \cref{alg:modified_new}.}
\label{alg:new_optimized}
\begin{algorithmic}
\Require $(\symInputElement_0,\symInputElement_1,\ldots,\symInputElement_{\symInputSize-1})$
\Ensure $(
\symHashElement_0,
\symHashElement_1,
\ldots,
\symHashElement_{\symHashSize-1}
)\in [0, \symHashSize)^\symHashSize$
\State $(
\symHashElement_0,
\symHashElement_1,
\ldots,
\symHashElement_{\symHashSize-1}
)\gets
(\infty,\infty,\ldots,\infty)$
\State allocate array $(\symShuffleArrayPos_0,\symShuffleArrayPos_1,\dots,\symShuffleArrayPos_{\symHashSize-1})$
\State $(\symShuffleArrayStamp_0,\symShuffleArrayStamp_1,\dots,\symShuffleArrayStamp_{\symHashSize-1})\gets (-1,-1,\ldots,-1)$
\State $(\symHistogram_0,\symHistogram_1,\ldots,\symHistogram_{\symHashSize-2},\symHistogram_{\symHashSize-1}) \gets (0,0,\ldots, 0, \symHashSize)$
\State $\symMaxNonZero \gets \symHashSize-1$
\For{$\symInputIndex \gets 0,1,\ldots, \symInputSize-1$}
\State initialize pseudo-random generator with seed $\symInputElement_\symInputIndex$
\State $\symHashIndex \gets 0$
\While{$\symHashIndex \leq \symMaxNonZero$}
\State $\symIntermediateHash\gets$ uniform random number from $[0,1)$
\State $\symShuffleIndex \gets$ uniform random  number from $\lbrace \symHashIndex, \ldots, \symHashSize-1\rbrace$
\If{$\symShuffleArrayStamp_\symHashIndex\neq \symInputIndex$}
\State $\symShuffleArrayStamp_\symHashIndex\gets\symInputIndex$
\State $\symShuffleArrayPos_\symHashIndex\gets\symHashIndex$
\EndIf
\If{$\symShuffleArrayStamp_\symShuffleIndex\neq \symInputIndex$}
\State $\symShuffleArrayStamp_\symShuffleIndex\gets\symInputIndex$
\State $\symShuffleArrayPos_\symShuffleIndex\gets\symShuffleIndex$
\EndIf
\State swap $\symShuffleArrayPos_\symHashIndex$ and $\symShuffleArrayPos_\symShuffleIndex$
\If{$\symIntermediateHash + \symHashIndex < \symHashElement_{\symShuffleArrayPos_\symHashIndex}$}
\State $\symHashIndex' \gets \min(\lfloor\symHashElement_{\symShuffleArrayPos_\symHashIndex} \rfloor, \symHashSize-1)$
\State $\symHashElement_{\symShuffleArrayPos_\symHashIndex} \gets \symIntermediateHash + \symHashIndex$
\If{$\symHashIndex < \symHashIndex'$}
\State $\symHistogram_{\symHashIndex'}\gets \symHistogram_{\symHashIndex'} -1$
\State $\symHistogram_{\symHashIndex}\gets \symHistogram_{\symHashIndex} + 1$
\While{$\symHistogram_{\symMaxNonZero} = 0$}
\State $\symMaxNonZero \gets \symMaxNonZero - 1$
\EndWhile
\EndIf
\EndIf
\State $\symHashIndex\gets\symHashIndex+1$
\EndWhile
\EndFor
\end{algorithmic}
\end{algorithm}

As first step towards our final algorithm we merge both inner loops in \cref{alg:plain_new} and eliminate the initialization of array $(\symShuffleArrayPos_0,\symShuffleArrayPos_1,\ldots, \symShuffleArrayPos_{\symHashSize-1})$ as demonstrated by \cref{alg:modified_new}. The trick is to introduce a second array $(\symShuffleArrayStamp_0,\symShuffleArrayStamp_1,\ldots, \symShuffleArrayStamp_{\symHashSize-1})$ which is used to mark corresponding entries in $(\symShuffleArrayPos_0,\symShuffleArrayPos_1,\ldots, \symShuffleArrayPos_{\symHashSize-1})$ as initialized during the $\symHashIndex$-th inner loop cycle. $\symShuffleArrayPos_\symShuffleIndex$ is regarded as initialized if and only if $\symShuffleArrayStamp_\symShuffleIndex = \symHashIndex$. Otherwise, $\symShuffleArrayPos_\symShuffleIndex$ is set equal to $\symShuffleIndex$ when accessed first and $\symShuffleArrayStamp_\symShuffleIndex$ is simultaneously set equal to $\symHashIndex$ to flag the entry as initialized.

A second modification compared to \cref{alg:plain_new} is that the signature value update $\symHashElement_\symHashIndex \gets \min(\symHashElement_\symHashIndex, \symIntermediateHash + \symShuffleArrayPos_\symHashIndex)$ has been replaced by
$\symHashElement_{\symShuffleArrayPos_\symHashIndex} \gets \min(\symHashElement_{\symShuffleArrayPos_\symHashIndex}, \symIntermediateHash + \symHashIndex)$. Both variants are statistically equivalent, because it does not make any difference, whether we interpret the randomly generated permutation as $\symPermutation(\symInputElement)$ or as its inverse $\symPermutation^{-1}(\symInputElement)$.

\cref{alg:modified_new} shows potential for further improvement. We see that the signature value updates $\symIntermediateHash + \symHashIndex$ are strictly increasing within the inner loop. Therefore, if we knew the current maximum of all current signature values, we would be able to leave the inner loop early.
The solution is to  maintain a histogram over the integral parts of the current signature values
\begin{equation*}
\symHistogram_\symHashIndexTwo := 
\begin{cases}
\sum_{\symHashIndex=0}^{\symHashSize-1} \symIndicator(\lfloor\symHashElement_\symHashIndex\rfloor = \symHashIndexTwo)
& 
\symHashIndexTwo \in \lbrace 0,1,\ldots, \symHashSize-2\rbrace
\\
\sum_{\symHashIndex=0}^{\symHashSize-1} \symIndicator(\symHashElement_\symHashIndex \geq \symHashSize-1)
& 
\symHashIndexTwo = \symHashSize-1
\end{cases}
\end{equation*}
and also to keep track of the maximum non-zero histogram entry
\begin{equation*}
\symMaxNonZero := \max (\lbrace\symHashIndex \mid \symHistogram_\symHashIndex > 0\rbrace).
\end{equation*}
Knowing $\symMaxNonZero$ allows escaping the inner loop as soon as $\symHashIndex > \symMaxNonZero$, because further signature value updates are not possible in this case.  The result of all these optimizations is the new SuperMinHash algorithm as shown in
\cref{alg:new_optimized}.

\subsection{Precision}

\begin{figure}
\centering
\includegraphics[width=1\columnwidth]{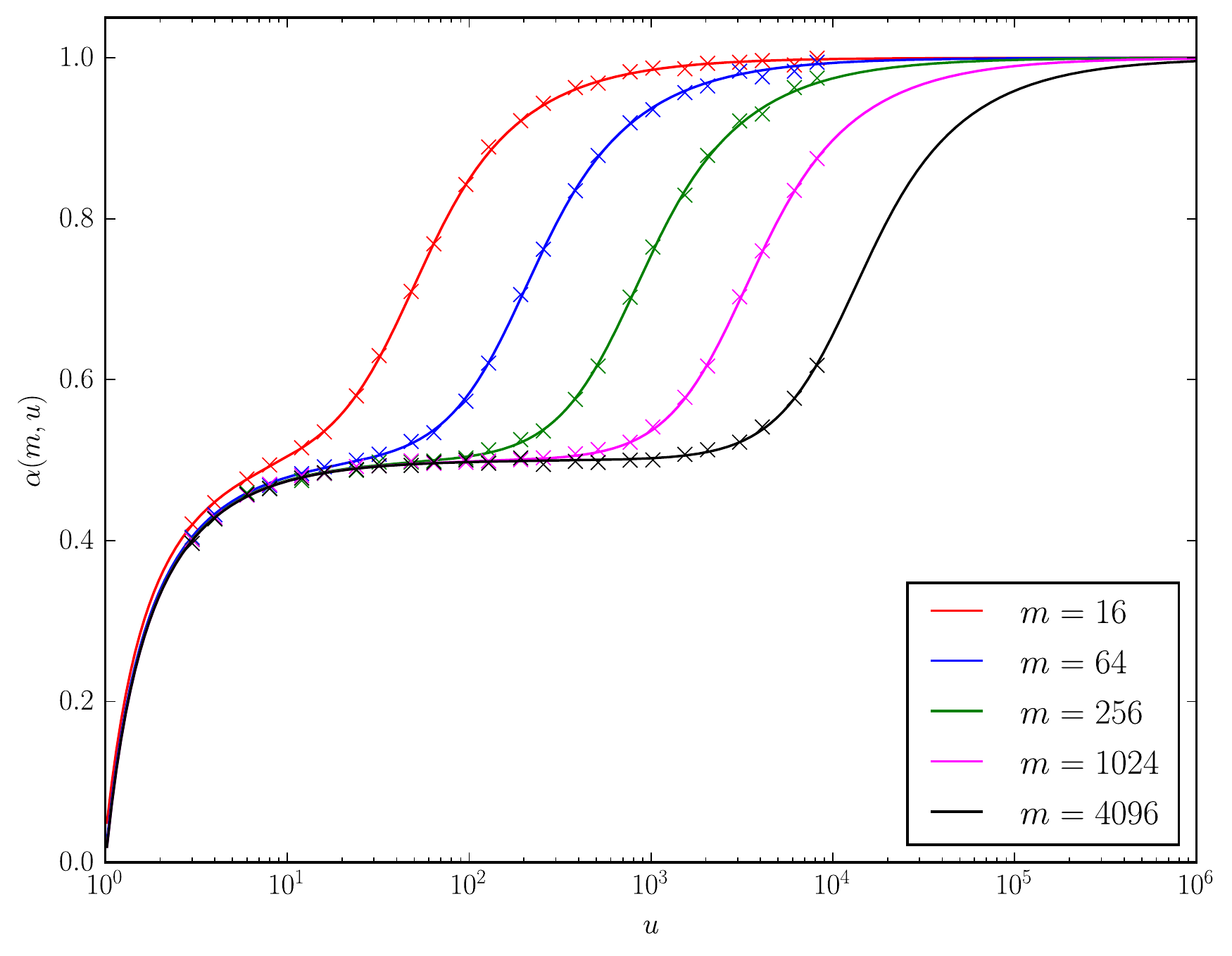}
\caption{The function $\symImprovement(\symHashSize,\symUnionSize)$ over $\symUnionSize$ for different signature sizes $\symHashSize$. The crosses represent values obtained through simulations.}
\label{fig:variance}
\end{figure}

As proven in the appendix the variance of estimator \eqref{equ:jaccard_estimator} for the new signature is
\begin{equation}
\label{equ:variance_new}
\symVariance(\symJaccardEstimator)
=
\frac{\symJaccard(1-\symJaccard)}{\symHashSize}
\alpha(\symHashSize,\symUnionSize)
\end{equation}
where $\symUnionSize := |\symSetA \cup \symSetB|$ is the union cardinality. The function $\symImprovement(\symHashSize,\symUnionSize)$ is defined as
\begin{equation}
\label{equ:improvement_factor}
\symImprovement(\symHashSize,\symUnionSize)
:=
1
-
\frac{\sum_{\symSumIndex=1}^{\symHashSize-1}
\symSumIndex^{\symUnionSize}
\left(
(\symSumIndex + 1)^{\symUnionSize}
+
(\symSumIndex - 1)^{\symUnionSize}
-
2
{\symSumIndex}^{\symUnionSize}
\right)}{(\symHashSize-1)^{\symUnionSize-1} \symHashSize^{\symUnionSize} (\symUnionSize-1)}.
\end{equation}
The function is always in the range $[0, 1)$, because the term $
(\symSumIndex + 1)^{\symUnionSize}
+
(\symSumIndex - 1)^{\symUnionSize}
-
2
{\symSumIndex}^{\symUnionSize}
$ is positive for $\symUnionSize > 1$. $\symImprovement(\symHashSize,\symUnionSize)$ corresponds to the reduction factor of the variance relative to that of MinHash signatures \eqref{equ:minwise_hashing_variance}. \cref{fig:variance} shows the function for different values of $\symHashSize$. Interestingly, $\symImprovement(\symHashSize,\symUnionSize)$ only depends on the union cardinality $\symUnionSize$ and the signature size $\symHashSize$ and does not depend on the Jaccard index $\symJaccard$. Compared to MinHash the variance is approximately by a factor of two smaller in case $\symUnionSize < \symHashSize$.

To verify \eqref{equ:variance_new} we have conducted some simulations to determine the variance of the Jaccard index estimator for two random sets $\symSetA$ and $\symSetB$ experimentally. We considered the cases $|\symSetA\setminus\symSetB| = |\symSetB\setminus\symSetA| = |\symSetA\cap\symSetB| = 2^k$ with $\symUnionSize = 3\cdot2^k$ and the cases $|\symSetA\setminus\symSetB|/2 = |\symSetB\setminus\symSetA| = |\symSetA\cap\symSetB| = 2^k$ with $\symUnionSize = 4\cdot2^k$ both for $k\in\lbrace{0, 1,\ldots, 11\rbrace}$. For each case \num{100000} different triples of disjoint sets $\symSetS_{\symSetA\setminus\symSetB}$, $\symSetS_{\symSetB\setminus\symSetA}$, and $\symSetS_{\symSetB\cap\symSetA}$ have been randomly generated with cardinalities 
$|\symSetA\setminus\symSetB|$, $|\symSetB\setminus\symSetA|$, and $|\symSetA\cap\symSetB|$, respectively. Then the sets $\symSetA$ and $\symSetB$ are constructed using $\symSetA = \symSetS_{\symSetA\setminus\symSetB} \cup \symSetS_{\symSetA\cap\symSetB}$ and $\symSetB = \symSetS_{\symSetB\setminus\symSetA} \cup \symSetS_{\symSetA\cap\symSetB}$. After calculating the corresponding hash signatures, their common Jaccard index has been estimated. The estimates of all \num{100000} simulation runs have been used to calculate the variance and also $\symImprovement(\symHashSize,\symUnionSize)$ by dividing by the theoretical MinHash variance \eqref{equ:minwise_hashing_variance}. The experimental results are shown as crosses in \cref{fig:variance} and confirm the theoretically derived formula \eqref{equ:improvement_factor}. 

For all simulation runs we used the 128-bit version of the MurmurHash3 algorithm which also allows to specify a seed. We used a predefined sequence of seed values to generate an arbitrary number of hash values for a given data element, which are used as bit source for pseudo-random number generation.
\subsection{Runtime}

To analyze the runtime of \cref{alg:new_optimized} we first consider the case that all inserted elements are distinct ($\symInputSize = |\symInputSet|$). The expected runtime is given by the expected total number of inner (while) loop iterations denoted by $\symRunningTime = \symRunningTime(\symInputSize, \symHashSize)$ that are needed when inserting $\symInputSize$ elements. If $\symRuntime_{\symMaxNonZeroValue}$ denotes the average number of element insertions until $\symMaxNonZero$ becomes smaller than $\symMaxNonZeroValue$, we can write
\begin{equation*}
\symRunningTime(\symInputSize, \symHashSize)
=
\symInputSize + \sum_{\symMaxNonZeroValue = 1}^{\symHashSize-1}\min(\symRuntime_\symMaxNonZeroValue,\symInputSize).
\end{equation*}
Since $\symMaxNonZero$ is smaller than $\symMaxNonZeroValue$ as soon as each signature value is less than $\symMaxNonZeroValue$, $\symRuntime_\symMaxNonZeroValue$ can be regarded as the average number of random permutations that are necessary until any value of $\lbrace 0, 1, \ldots,\symMaxNonZeroValue-1 \rbrace$ was mapped to each signature index. This corresponds to the coupon collector's problem with collection size $\symHashSize$ and group drawings of size $\symMaxNonZeroValue$, where each drawing gives $\symMaxNonZeroValue$ distinct coupons \cite{Stadje1990}.
In our case the complete collection corresponds to the $\symHashSize$ signature indices. Drawing a group of coupons corresponds to selecting the first $\symMaxNonZeroValue$ indices after permuting a list with all $\symHashSize$ of them.

For the classical coupon  collector's problem with group size $\symMaxNonZeroValue = 1$ we have the well known solution \cite{Cormen2009}
\begin{equation*}
\symRuntime_1 = \symHashSize \symHarmonicNumber_\symHashSize.
\end{equation*}
Here $\symHarmonicNumber_\symHashSize := \frac{1}{1} + \frac{1}{2} + \ldots +\frac{1}{\symHashSize}$ denotes the $\symHashSize$-th harmonic number. Unfortunately, there is no simple expression for $\symMaxNonZeroValue \geq 2$ \cite{Stadje1990}. However, it is easy to find an upper bound for $\symRuntime_\symMaxNonZeroValue$. Let $\symProbabilityMass_\symSumIndex$ be the probability that $\symSumIndex$ drawings are necessary to complete the coupon collection for the classical case with group size 1. By definition, we have $\sum_{\symSumIndex=1}^\infty \symProbabilityMass_\symSumIndex \symSumIndex = \symRuntime_1 = \symHashSize \symHarmonicNumber_\symHashSize$ with 
$\sum_{\symSumIndex=1}^\infty \symProbabilityMass_\symSumIndex  =1$.
If $\symSumIndex$ drawings are necessary to complete the collection for the case $\symMaxNonZeroValue = 1$, it is obvious that not more than $\lceil \symSumIndex / \symMaxNonZeroValue\rceil$ drawings will be necessary for the general case with group size $\symMaxNonZeroValue$. Therefore, we can find the upper bound
\begin{equation*}
\symRuntime_\symMaxNonZeroValue 
\leq
\sum_{\symSumIndex=1}^\infty \symProbabilityMass_\symSumIndex \left\lceil \frac{\symSumIndex}{\symMaxNonZeroValue}\right\rceil
\leq
\sum_{\symSumIndex=1}^\infty \symProbabilityMass_\symSumIndex \frac{\symSumIndex+\symMaxNonZeroValue - 1}{\symMaxNonZeroValue}
=
\frac{\symHashSize \symHarmonicNumber_\symHashSize + \symMaxNonZeroValue - 1}{\symMaxNonZeroValue}.
\end{equation*}
Using this inequality together with $\min(\symRuntime_\symMaxNonZeroValue,\symInputSize) \leq \symRuntime_\symMaxNonZeroValue$ we get
\begin{align}
\symRunningTime(\symInputSize, \symHashSize)
&\leq
\symInputSize + \sum_{\symMaxNonZeroValue = 1}^{\symHashSize-1}\symRuntime_\symMaxNonZeroValue
\leq
\symInputSize
+
(\symHashSize \symHarmonicNumber_{\symHashSize}-1) \symHarmonicNumber_{\symHashSize-1} + \symHashSize-1
\nonumber
\\
\label{equ:perf_bound_1}
&=
\symInputSize + \symBigO(\symHashSize\,\log^2\symHashSize)
=
\symBigO(\symInputSize + \symHashSize\,\log^2\symHashSize).
\end{align}
Here we used the relationship $\symHarmonicNumber_{\symHashSize}=\symBigO(\log\symHashSize)$. In any case the worst case runtime is limited by the maximum number of inner loop iterations, which is equal to $\symInputSize \symHashSize$, if the shortcut introduced in \cref{alg:new_optimized} never comes into play. Thus, the new algorithm never needs more inner loop cycles than the MinHash algorithm.

\begin{figure}
\centering
\includegraphics[width=1\columnwidth]{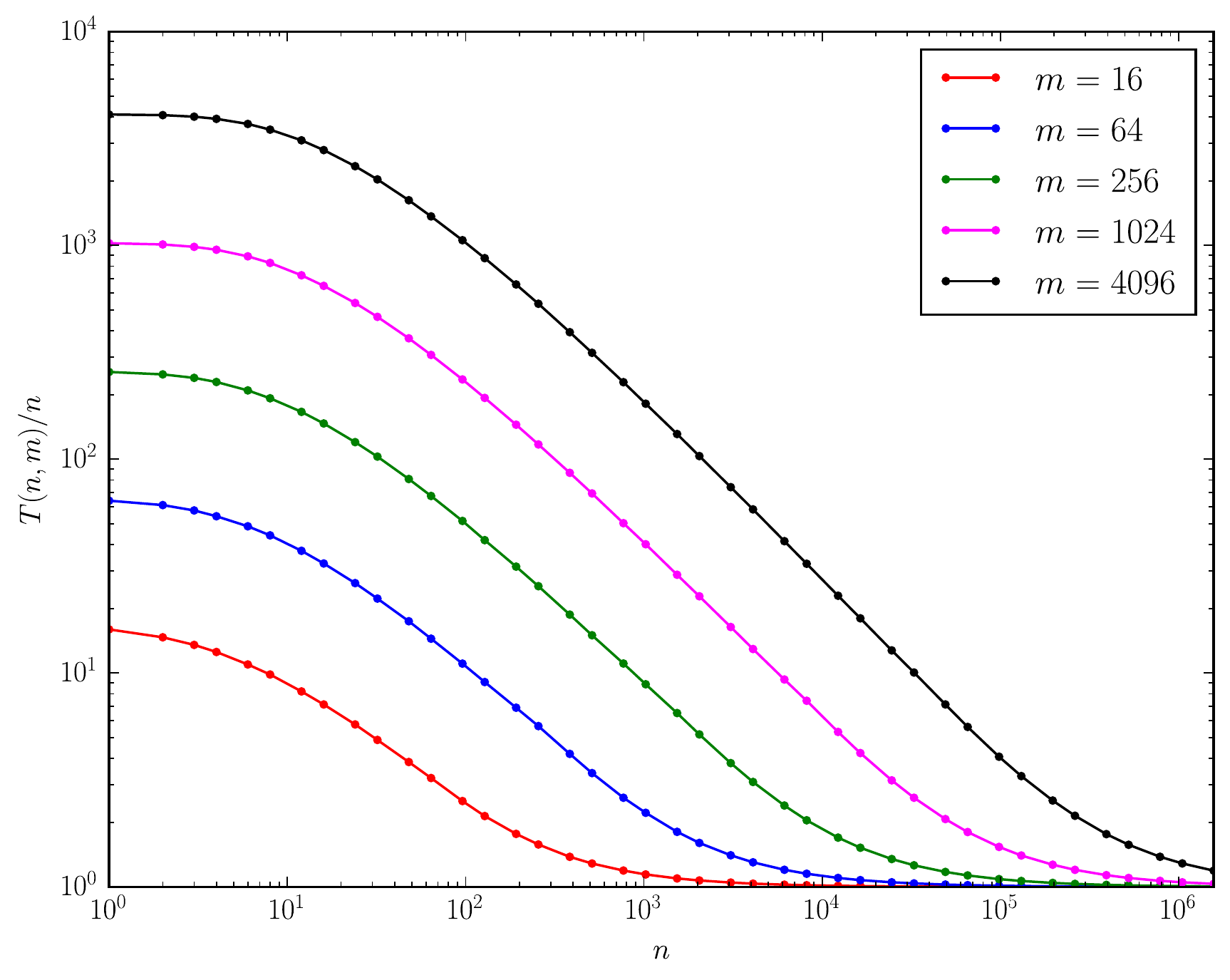}
\caption{The average number of inner loop iterations in \cref{alg:new_optimized} per inserted data element over the set size $\symInputSize$ for different signature sizes $\symHashSize$.}
\label{fig:performance}
\end{figure}

To better understand the runtime of \cref{alg:new_optimized} compared to the MinHash algorithm, we investigated the average number of inner loop cycles per inserted data element $\symRunningTime(\symInputSize, \symHashSize)/\symInputSize$. For the new algorithm we expect that that this number starts at $\symHashSize$ and decreases to 1 as $\symInputSize\rightarrow\infty$ because of \eqref{equ:perf_bound_1}. 
In contrast, the MinHash algorithm always needs $\symHashSize$ inner loop iterations regardless of the input data size $\symInputSize$. 

\cref{fig:performance} shows the experimentally determined average number of inner loop cycles for set sizes $\symInputSize=2^k$ with $k\in\lbrace 0,1,\ldots, 20\rbrace$ and $\symInputSize=3\cdot 2^k$ with $k\in\lbrace 0,1,\ldots, 19\rbrace$ based on \num{1000} randomly generated data sets, respectively. The results indeed show that the amortized costs for a single data element insertion correspond to a single inner loop execution as $\symInputSize \rightarrow \infty$. Thus, the runtime performance is improved up to a factor of $\symHashSize$ relative to MinHash.

If the input data $\symInputElement_0, \symInputElement_1, \ldots, \symInputElement_{\symInputSize-1}$ contains duplicates, hence $|\symInputSet| < \symInputSize$, the runtime will be longer, because repeated insertions of identical values will not change any state in \cref{alg:new_optimized}. If we assume that the order of input elements is entirely random, the 
average number of inner loop iterations will be $\symRunningTime(|\symInputSet|, \symHashSize) \frac{\symInputSize}{|\symInputSet|}$ which gives $\symRunningTime(|\symInputSet|, \symHashSize)/|\symInputSet|$ per inserted element. In the worst case, if many input elements are equal and sorted, the number of inner loop iterations per input element is still limited by $\symHashSize$ and thus equal to that of MinHash.

\section{Conclusion}
We have presented the SuperMinHash algorithm which can be used as full replacement for the MinHash algorithm as it has similar or even better properties as the MinHash algorithm. The new  algorithm has comparable runtime for small input sizes, is significantly faster for larger data sizes, can also be used as streaming algorithm as it requires only a single pass over the data, and significantly improves the precision of Jaccard index estimates for small sets.
\appendix
\section*{Appendix}
\label{sec:appendix}

In order to derive formula \eqref{equ:variance_new} for the variance of estimator  \eqref{equ:jaccard_estimator} when applied to the new signature we first consider the conditional probability 
$\symProbability(\symHashElement_\symHashIndex(\symSetA) = \symHashElement_\symHashIndex(\symSetB) \mid \symHashElement_\symHashIndexTwo(\symSetA) = \symHashElement_\symHashIndexTwo(\symSetB))$. For the trivial case $\symHashIndex=\symHashIndexTwo$ this probability is equal to 1.
Therefore, we consider the case $\symHashIndex\neq\symHashIndexTwo$ in the following.

If we introduce 
\begin{equation}
\label{equ:input_min}
\symInputElementMin := \argmin_{\symInputElement \in \symSetA \cup \symSetB} \symHashElement_\symHashIndexTwo(\symInputElement)
\end{equation}
and use the equivalences $\symHashElement_\symHashIndex(\symSetA) = \symHashElement_\symHashIndex(\symSetB) \Leftrightarrow \symHashElement_\symHashIndex(\symSetA\cap\symSetB)
=
\symHashElement_\symHashIndex(\symSetA\cup\symSetB)
\Leftrightarrow
\symHashElement_\symHashIndex(\symSetA\cap\symSetB) < \symHashElement_\symHashIndex(\symSetA\bigtriangleup\symSetB)
$
and
$
\symHashElement_\symHashIndexTwo(\symSetA) = \symHashElement_\symHashIndexTwo(\symSetB)
\Leftrightarrow
\symInputElementMin \in \symSetA \cap \symSetB
$,
where $\symSetA\bigtriangleup\symSetB$ denotes the symmetric difference of sets $\symSetA$ and $\symSetB$,
we can write
\begin{align}
\symProbability(\symHashElement_\symHashIndex(\symSetA) &= \symHashElement_\symHashIndex(\symSetB)
\mid
\symHashElement_\symHashIndexTwo(\symSetA) = \symHashElement_\symHashIndexTwo(\symSetB)
)
\nonumber
\\
&=
\symProbability(\symHashElement_\symHashIndex(\symSetA\cap\symSetB) < \symHashElement_\symHashIndex(\symSetA\bigtriangleup\symSetB)
\mid
\symInputElementMin \in \symSetA \cap \symSetB)
\nonumber
\\
\label{equ:probability_equal}
&=
\int\limits_0^\symHashSize
R'(z) S(z)   dz
\end{align}
with functions $R(z)$ and $S(z)$
defined as 
\begin{equation*}
R(z) := \symProbability(\symHashElement_\symHashIndexTwo(\symInputElementMin)< z
\mid
\symInputElementMin \in \symSetA \cap \symSetB)
=
\symProbability(\symHashElement_\symHashIndexTwo(\symInputElementMin)< z)
\end{equation*}
and
\begin{equation*}
S(z) :=
\symProbability(\symHashElement_\symHashIndex(\symSetA\cap\symSetB) < \symHashElement_\symHashIndex(\symSetA\bigtriangleup\symSetB)
\mid
\symInputElementMin \in \symSetA \cap \symSetB
\wedge
\symHashElement_\symHashIndexTwo(\symInputElementMin)=z),
\end{equation*}
respectively. 

Since  $\symHashElement_\symHashIndexTwo(\symInputElementMin)$ is the minimum of $\symUnionSize := |\symSetA\cup\symSetB|$ independent uniformly distributed values from $[0, \symHashSize)$, its cumulative distribution function is
\begin{equation*}
R(z)
=
1 - \left(1 - \frac{z}{\symHashSize}\right)^{\symUnionSize}
\end{equation*}
and its first derivative is
\begin{equation}
\label{equ:func_r_deriv}
R'(z)
=
\frac{\symUnionSize}{\symHashSize^\symUnionSize}\left(\symHashSize - z\right)^{\symUnionSize-1}.
\end{equation}

To get $S(z)$ we first consider the distribution of 
$\symHashElement_\symHashIndex(\symInputElement)$ for any input element $\symInputElement$ conditioned on $\symHashElement_\symHashIndexTwo(\symInputElement)=z$. The distribution of $\symHashElement_\symHashIndex(\symInputElement)$ is uniform over $[0,\lfloor z \rfloor) \cup [\lfloor z \rfloor + 1, \symHashSize)$, because the integral part must be different from $\lfloor z \rfloor$ due to the permutation in \eqref{equ:def_new_hash}. The corresponding complementary cumulative distribution function is 
\begin{align}
V(x,z) &:=
\symProbability(
		\symHashElement_\symHashIndex(\symInputElement)
> x 
	\mid 
		\symHashElement_\symHashIndexTwo(
			\symInputElement)
		= z)
\nonumber
\\
\label{equ:def_v}
&=
\frac{1}{\symHashSize-1}\cdot
\begin{cases}
\symHashSize - 1 - x 
&
x \in [0, \lfloor z \rfloor)
\\
\symHashSize - 1 - \lfloor z \rfloor
&
x \in [\lfloor z \rfloor, \lfloor z \rfloor + 1)
\\
\symHashSize - x 
&
x \in [\lfloor z \rfloor + 1, \symHashSize).
\end{cases}		
\end{align}
Next we consider the distribution of $\symHashElement_\symHashIndex(\symInputElement)$, if $\symInputElement \neq \symInputElementMin$ and $\symHashElement_\symHashIndexTwo(\symInputElementMin) = z$. 
Due to \eqref{equ:input_min} $\symInputElement \neq \symInputElementMin$ is equivalent to $\symHashElement_\symHashIndexTwo(\symInputElement)>\symHashElement_\symHashIndexTwo(\symInputElementMin)$. Furthermore,
$\symHashElement_\symHashIndexTwo(\symInputElement)$ is uniformly distributed over $[z, \symHashSize)$. Therefore, we get for the complementary cumulative distribution function
\begin{align}
W(x,z)
&:=
\symProbability(
		\symHashElement_\symHashIndex(\symInputElement)
> x 
	\mid 
	\symInputElement\neq\symInputElementMin
	\wedge
		\symHashElement_\symHashIndexTwo(
			\symInputElementMin)
		= z
		)
\nonumber\\
&=
\symProbability(
		\symHashElement_\symHashIndex(\symInputElement)
> x 
	\mid 
	\symHashElement_\symHashIndexTwo(\symInputElement)> 
	\symHashElement_\symHashIndexTwo(\symInputElementMin)
\wedge
	\symHashElement_\symHashIndexTwo(\symInputElementMin) = z
		)
\nonumber\\
&=
\symProbability(
		\symHashElement_\symHashIndex(\symInputElement)
> x 
	\mid 
	\symHashElement_\symHashIndexTwo(\symInputElement)> z
\wedge
	\symHashElement_\symHashIndexTwo(\symInputElementMin) = z
		)
\nonumber\\
&=
\symProbability(
		\symHashElement_\symHashIndex(\symInputElement)
> x 
	\mid 
	\symHashElement_\symHashIndexTwo(\symInputElement)> z
		)
\nonumber\\
&=
\frac{1}{\symHashSize - z}
\int\limits_{z}^{\symHashSize}
\symProbability(
		\symHashElement_\symHashIndex(\symInputElement)
		> x 
	\mid 
		\symHashElement_\symHashIndexTwo(
			\symInputElement)
		= z') dz'
\nonumber\\
&=
\frac{1}{\symHashSize - z}\int\limits_{z}^{\symHashSize}
V(x,z') dz'
\nonumber\\
\label{equ:def_w}
&=
\begin{cases}
\frac{\symHashSize-1-x}{\symHashSize-1} & 
x \in [0, \lfloor z \rfloor)
\\
\frac{
(\symHashSize - 1 - \lfloor z \rfloor)
(\symHashSize - (x - \lfloor z \rfloor) - z)
}{
(\symHashSize-1)
(\symHashSize - z)
}
&
x \in [\lfloor z \rfloor, \lfloor z \rfloor + 1)
\\
\frac{
(\symHashSize - x)
(\symHashSize - 1 - z)
}{
(\symHashSize-1)
(\symHashSize - z)
}
&
x \in [\lfloor z \rfloor + 1, \symHashSize).
\end{cases}
\end{align}

Now we are able to determine the complementary cumulative distribution functions for $\symHashElement_\symHashIndex(\symSetA \bigtriangleup \symSetB)$, which is the minimum of $|\symSetA \bigtriangleup \symSetB|=\symUnionSize(1-\symJaccard)$ identically distributed random variables obeying \eqref{equ:def_w}, conditioned on $\symInputElementMin\in\symSetA \cap \symSetB$ and $\symHashElement_\symHashIndexTwo(\symInputElementMin)
= z$ 
\begin{align}
&\symProbability(\symHashElement_\symHashIndex(\symSetA \bigtriangleup \symSetB) 
>
x
\mid 
\symInputElementMin\in\symSetA \cap \symSetB
\wedge
\symHashElement_\symHashIndexTwo(\symInputElementMin)
= z) 
\nonumber
\\
&\qquad=
\left(\symProbability(
		\symHashElement_\symHashIndex(\symInputElement)
> x 
	\mid 
\symInputElement \neq \symInputElementMin
	\wedge
		\symHashElement_\symHashIndexTwo(
			\symInputElementMin)
		= z
		)
\right)^{|\symSetA \bigtriangleup \symSetB|}
\nonumber
\\
\label{equ:prob_delta}
&\qquad=
\left(W(x,z)\right)^{\symUnionSize(1-\symJaccard)}.
\end{align}
Here we have used the fact that the complementary cumulative distribution function of the minimum of independent random variables is identical to the product of the individual complementary cumulative distribution functions.

Analogously, $\symHashElement_\symHashIndex(\symSetA \cap \symSetB)$ conditioned on $\symInputElementMin\in\symSetA \cap \symSetB$ and $\symHashElement_\symHashIndexTwo(\symInputElementMin)=z$ is distributed like the minimum of $|\symSetA \cap \symSetB|-1 = \symUnionSize\symJaccard-1$ identically distributed random variables following \eqref{equ:def_w} and $\symHashElement_\symHashIndex(\symInputElementMin)$ which is described by \eqref{equ:def_v}
\begin{align}
&\symProbability(\symHashElement_\symHashIndex(\symSetA \cap \symSetB) 
>
x
\mid 
\symInputElementMin\in\symSetA \cap \symSetB
\wedge
\symHashElement_\symHashIndexTwo(\symInputElementMin)
=
z)
\nonumber
\\
&\qquad=
\symProbability(
		\symHashElement_\symHashIndex(\symInputElement')
> x 
	\mid 
		\symHashElement_\symHashIndexTwo(
			\symInputElementMin)
		= z
		)
\cdot
\nonumber
\\
&\qquad\qquad\cdot
\left(\symProbability(
		\symHashElement_\symHashIndex(\symInputElement)
> x 
	\mid 
\symInputElement \neq \symInputElementMin
	\wedge
		\symHashElement_\symHashIndexTwo(
			\symInputElementMin)
		= z
		)
\right)^{|(\symSetA \cap \symSetB) \setminus \lbrace \symInputElementMin\rbrace|}
\nonumber
\\
\label{equ:prob_cap}
&\qquad=
V(x,z)\left(W(x,z)\right)^{\symUnionSize\symJaccard-1}
.
\end{align}
Using \eqref{equ:prob_delta} and \eqref{equ:prob_cap} we can derive $S(z)$

\begin{align}
&
S(z)
=
\symProbability(\symHashElement_\symHashIndex(\symSetA\cap\symSetB) < \symHashElement_\symHashIndex(\symSetA\bigtriangleup\symSetB)
\mid 
\symInputElementMin\in\symSetA \cap \symSetB
\wedge
\symHashElement_\symHashIndexTwo(\symInputElementMin)=z)
\nonumber\\
&\quad=
1-
\symProbability(\symHashElement_\symHashIndex(\symSetA\cap\symSetB) > \symHashElement_\symHashIndex(\symSetA\bigtriangleup\symSetB)
\mid
\symInputElementMin\in\symSetA \cap \symSetB
\wedge
\symHashElement_\symHashIndexTwo(\symInputElementMin)=z)
\nonumber\\
&\quad=
1+
\int\limits_{0}^\symHashSize
\left(
\begin{aligned}
&\symProbability(\symHashElement_\symHashIndex(\symSetA \cap \symSetB) 
>
x
\mid 
\symInputElementMin\in\symSetA \cap \symSetB
\wedge
\symHashElement_\symHashIndexTwo(\symInputElementMin)
= z)
\cdot
\\
&
\cdot
\frac{\partial\symProbability(\symHashElement_\symHashIndex(\symSetA \bigtriangleup \symSetB) 
>
x
\mid 
\symInputElementMin\in\symSetA \cap \symSetB
\wedge
\symHashElement_\symHashIndexTwo(\symInputElementMin)
= z) 
}{\partial x} 
\end{aligned}
\right)
dx
\nonumber\\
&\quad=
1+\symUnionSize(1-\symJaccard)
\int\limits_{0}^\symHashSize
V(x,z)
\left(W(x,z)\right)^{\symUnionSize-2}
\frac{\partial W(x,z)}{\partial x}
dx
\nonumber\\
&\quad=
1+\frac{\symUnionSize(1-\symJaccard)}
{\symUnionSize-1}
\int\limits_{0}^\symHashSize
{V}(x,z)
\frac{\partial}{\partial x}\left(\left({W}(x,z)\right)^{\symUnionSize-1}\right)
dx
\nonumber\\
&\quad=
1-\frac{\symUnionSize(1-\symJaccard)}
{\symUnionSize-1}
\left(
1
+
\int\limits_{0}^\symHashSize
\frac{\partial {V}(x,z)}{\partial x}
\left({W}(x,z)\right)^{\symUnionSize-1}
dx
\right)
\nonumber\\
&\quad=
1-{\textstyle\frac{\symUnionSize(1-\symJaccard)}
{\symUnionSize-1}}
\left(
\begin{aligned}
&1
-
{\textstyle\frac{1}{\symHashSize-1}}
\int\limits_{0}^{\lfloor z \rfloor}
\left({\textstyle\frac{\symHashSize - 1 - x}{\symHashSize-1}}\right)^{\symUnionSize-1}
dx
\\
&
-
{\textstyle\frac{1}{\symHashSize-1}}
\int\limits_{\lfloor z \rfloor+1}^{\symHashSize}
\left({\textstyle\frac{
(\symHashSize - x)
(\symHashSize - 1 - z)
}{
(\symHashSize-1)
(\symHashSize - z)
}}
\right)^{\symUnionSize-1}
dx
\end{aligned}
\right)
\nonumber\\
&\quad=
1-{\textstyle\frac{\symUnionSize(1-\symJaccard)}
{\symUnionSize-1}}
\left(
\begin{aligned}
&
1
+
{\textstyle\frac{1}{\symUnionSize}}
\left[
\left({\textstyle\frac{\symHashSize - 1 - x}{\symHashSize-1}}\right)^{\symUnionSize}
\right]_{x = 0}^{\lfloor z \rfloor}
\\
&
+
{\textstyle
\frac{1}{\symUnionSize}
\left(\frac{\symHashSize - 1 - z}{\symHashSize - z}\right)^{\symUnionSize-1}
}
\left[
\left({\textstyle\frac{
\symHashSize - x
}{
\symHashSize-1
}}
\right)^{\symUnionSize}
\right]
_{x  = \lfloor z \rfloor+1}^{\symHashSize}
\end{aligned}
\right)
\nonumber\\
&\quad=
1-{\textstyle\frac{\symUnionSize(1-\symJaccard)}
{\symUnionSize-1}}
\left(
\begin{aligned}
&
1 + {\textstyle\frac{1}{\symUnionSize}}\left({\textstyle\frac{\symHashSize - 1 - \lfloor z \rfloor}{\symHashSize-1}}\right)^{\symUnionSize}-{\textstyle\frac{1}{\symUnionSize}}
\\
&
\quad
-
{\textstyle\frac{1}{\symUnionSize}}{\textstyle\left(\frac{\symHashSize - 1 - z}{\symHashSize - z}\right)^{\symUnionSize-1}}
{\textstyle\left(\frac{
\symHashSize -1- \lfloor z \rfloor
}{
\symHashSize-1
}
\right)^{\symUnionSize}
}
\end{aligned}
\right)
\nonumber\\
\label{equ:func_s}
&\quad=
\symJaccard
-
{\textstyle\frac{(1-\symJaccard)}
{\symUnionSize-1}}
\left({\textstyle\frac{\symHashSize - 1 - \lfloor z \rfloor}{\symHashSize-1}}\right)^{\symUnionSize}
\left(
1
-
\left({\textstyle\frac{
\symHashSize - 1 - z
}{
\symHashSize - z
}}
\right)^{\!\symUnionSize-1}
\right).
\end{align}

Now we can insert \eqref{equ:func_r_deriv} and \eqref{equ:func_s} into 
\eqref{equ:probability_equal} which gives

\begin{align}
&\symProbability(\symHashElement_\symHashIndex(\symSetA) = \symHashElement_\symHashIndex(\symSetB) \mid \symHashElement_\symHashIndexTwo(\symSetA) = \symHashElement_\symHashIndexTwo(\symSetB))
\nonumber\\
&\quad =
\int\limits_{0}^{\symHashSize} R'(z) S(z) dz
\nonumber\\
&\quad =
\symJaccard -
\int\limits_{0}^{\symHashSize} R'(z) \left(\symJaccard - S(z)\right) dz
\nonumber\\
&\quad =
\symJaccard
-
\sum_{\symSumIndex=0}^{\symHashSize-1}
\int\limits_{\symSumIndex}^{\symSumIndex+1} R'(z) \left(\symJaccard - S(z)\right) dz
\nonumber\\
&\quad =
\symJaccard
-
\frac{\symUnionSize}{\symHashSize^\symUnionSize}
\sum_{\symSumIndex=0}^{\symHashSize-1}
\int\limits_{\symSumIndex}^{\symSumIndex+1} 
\left(\symHashSize - z\right)^{\symUnionSize-1}
\left(\symJaccard - S(z)\right) dz
\nonumber\\
&\quad =
\symJaccard
-
{\textstyle\frac{\symUnionSize}{\symHashSize^\symUnionSize}}
{\textstyle\frac{1 - \symJaccard}
{\symUnionSize-1}}
\sum_{\symSumIndex=0}^{\symHashSize-1}
\int\limits_{\symSumIndex}^{\symSumIndex+1}
\left(
\begin{aligned}
\left(\symHashSize - z\right)^{\symUnionSize-1}
\left({\textstyle\frac{\symHashSize - 1 - \symSumIndex}{\symHashSize-1}}\right)^{\symUnionSize}
\cdot
\\
\cdot
\left(
1
-
\left({\textstyle\frac{
\symHashSize - 1 - z
}{
\symHashSize - z
}}
\right)^{\!\symUnionSize-1}
\right) 
\end{aligned}
\right)
dz
\nonumber\\
&\quad =
\symJaccard
-
{\textstyle\frac{\symUnionSize}{\symHashSize^\symUnionSize}}
{\textstyle\frac{1 - \symJaccard}
{\symUnionSize-1}}
\sum_{\symSumIndex=0}^{\symHashSize-2}
\left({\textstyle\frac{\symHashSize - 1 - \symSumIndex}{\symHashSize-1}}\right)^{\!\symUnionSize}
\int\limits_{\symSumIndex}^{\symSumIndex+1} 
\left(
\begin{aligned}
&\left(\symHashSize - z\right)^{\symUnionSize-1}
\\
&-
\left(
\symHashSize - 1 - z
\right)^{\symUnionSize-1}
\end{aligned}
\right)
dz
\nonumber\\
&\quad =
\symJaccard
-
{\textstyle
\frac{(1 - \symJaccard)
\sum_{\symSumIndex=0}^{\symHashSize-2}
\left(\symHashSize - 1 - \symSumIndex \right)^{\!\symUnionSize}
\left(
\left(\symHashSize - \symSumIndex \right)^{\!\symUnionSize}
+
\left( \symHashSize - 2 - \symSumIndex \right)^{\!\symUnionSize}
-
2
\left(\symHashSize - 1 - \symSumIndex\right)^{\!\symUnionSize}
\right)
}{(\symHashSize-1)^{\symUnionSize} \symHashSize^{\symUnionSize} (\symUnionSize-1)}}
\nonumber\\
&\quad =
\symJaccard
-
{\textstyle
\frac{(1 - \symJaccard)
\sum_{\symSumIndex=1}^{\symHashSize-1}
\symSumIndex^{\symUnionSize}
\left(
(\symSumIndex + 1)^{\symUnionSize}
+
(\symSumIndex - 1)^{\symUnionSize}
-
2
\symSumIndex^{\symUnionSize}
\right)
}{(\symHashSize-1)^{\symUnionSize} \symHashSize^{\symUnionSize} (\symUnionSize-1)}}
\nonumber\\
\label{equ:conditional_equal_prob}
&\quad =
\symJaccard
-
{\textstyle\frac{(1 - \symJaccard)(1 - \symImprovement(\symHashSize,\symUnionSize))}{\symHashSize-1}}.
\end{align}
Here we introduced $\symImprovement(\symHashSize,\symUnionSize)$ as defined in \eqref{equ:improvement_factor}.

To calculate the variance of the Jaccard index estimator we need the covariance of indicators $\symIndicator(\symHashElement_\symHashIndex(\symSetA) = \symHashElement_\symHashIndex(\symSetB))$ and $\symIndicator(\symHashElement_\symHashIndexTwo(\symSetA) = \symHashElement_\symHashIndexTwo(\symSetB))$
\begin{align*}
&\symCovariance(\symIndicator(\symHashElement_\symHashIndex(\symSetA) = \symHashElement_\symHashIndex(\symSetB)),\symIndicator(\symHashElement_\symHashIndexTwo(\symSetA) = \symHashElement_\symHashIndexTwo(\symSetB)))
\\
&\quad =
\symExpectation(\symIndicator(\symHashElement_\symHashIndex(\symSetA) = \symHashElement_\symHashIndex(\symSetB)) \symIndicator(\symHashElement_\symHashIndexTwo(\symSetA) = \symHashElement_\symHashIndexTwo(\symSetB))) 
\\
&\qquad \qquad
- 
\symExpectation(\symIndicator(\symHashElement_\symHashIndex(\symSetA) = \symHashElement_\symHashIndex(\symSetB)))
\symExpectation(\symIndicator(\symHashElement_\symHashIndexTwo(\symSetA) = \symHashElement_\symHashIndexTwo(\symSetB)))
\\
&\quad =
\symProbability(\symHashElement_\symHashIndex(\symSetA) = \symHashElement_\symHashIndex(\symSetB) \wedge \symHashElement_\symHashIndexTwo(\symSetA) = \symHashElement_\symHashIndexTwo(\symSetB)) 
\\
&\qquad \qquad
- 
\symProbability(\symHashElement_\symHashIndex(\symSetA) = \symHashElement_\symHashIndex(\symSetB))
\symProbability(\symHashElement_\symHashIndexTwo(\symSetA) = \symHashElement_\symHashIndexTwo(\symSetB))
\\
&\quad =
\symProbability(\symHashElement_\symHashIndex(\symSetA) = \symHashElement_\symHashIndex(\symSetB) \mid \symHashElement_\symHashIndexTwo(\symSetA) = \symHashElement_\symHashIndexTwo(\symSetB)) 
\,
\symProbability(
\symHashElement_\symHashIndexTwo(\symSetA) = \symHashElement_\symHashIndexTwo(\symSetB)
)
- \symJaccard^2
\\
&\quad =
\symProbability(\symHashElement_\symHashIndex(\symSetA) = \symHashElement_\symHashIndex(\symSetB) \mid \symHashElement_\symHashIndexTwo(\symSetA) = \symHashElement_\symHashIndexTwo(\symSetB))\,\symJaccard - \symJaccard^2
\\
&\quad =
\symJaccard\,(1- \symJaccard)\cdot
\begin{cases}
1 & \symHashIndex = \symHashIndexTwo \\
-\frac{1 - \alpha(\symHashSize,\symUnionSize)}{\symHashSize-1} & \symHashIndex \neq \symHashIndexTwo.
\end{cases}
\end{align*}
The last step used \eqref{equ:conditional_equal_prob} for the case $\symHashIndex \neq \symHashIndexTwo$.

Now we are finally able to derive the variance of the Jaccard index estimator \eqref{equ:jaccard_estimator}
\begin{align*}
\symVariance(\symJaccardEstimator) &=\frac{1}{\symHashSize^2} \symVariance\left(\sum\limits_{\symHashIndex = 0}^{\symHashSize-1}\symIndicator(\symHashElement_\symHashIndex(\symSetA) = \symHashElement_\symHashIndex(\symSetB))\right)
\\
&=
\frac{1}{\symHashSize^2}
\sum\limits_{\symHashIndex = 0}^{\symHashSize-1}
\sum\limits_{\symHashIndexTwo = 0}^{\symHashSize-1}
\symCovariance(\symIndicator(\symHashElement_\symHashIndex(\symSetA) = \symHashElement_\symHashIndex(\symSetB)),\symIndicator(\symHashElement_\symHashIndexTwo(\symSetA) = \symHashElement_\symHashIndexTwo(\symSetB)))
\\
&=
\frac{\symJaccard(1-\symJaccard)}{\symHashSize^2}
\left(
\symHashSize
-
\symHashSize\,(\symHashSize-1)\frac{1 - \alpha(\symHashSize,\symUnionSize)}{\symHashSize-1}
\right)
\\
&=
\frac{\symJaccard(1-\symJaccard)}{\symHashSize}
\alpha(\symHashSize,\symUnionSize).
\end{align*}

\bibliographystyle{ACM-Reference-Format}

\bibliography{bibliography.bib}


\begin{thebibliography}{00}


\ifx \showCODEN    \undefined \def \showCODEN     #1{\unskip}     \fi
\ifx \showDOI      \undefined \def \showDOI       #1{#1}\fi
\ifx \showISBNx    \undefined \def \showISBNx     #1{\unskip}     \fi
\ifx \showISBNxiii \undefined \def \showISBNxiii  #1{\unskip}     \fi
\ifx \showISSN     \undefined \def \showISSN      #1{\unskip}     \fi
\ifx \showLCCN     \undefined \def \showLCCN      #1{\unskip}     \fi
\ifx \shownote     \undefined \def \shownote      #1{#1}          \fi
\ifx \showarticletitle \undefined \def \showarticletitle #1{#1}   \fi
\ifx \showURL      \undefined \def \showURL       {\relax}        \fi
\providecommand\bibfield[2]{#2}
\providecommand\bibinfo[2]{#2}
\providecommand\natexlab[1]{#1}
\providecommand\showeprint[2][]{arXiv:#2}

\bibitem[\protect\citeauthoryear{Broder}{Broder}{1997}]%
        {Broder1997}
\bibfield{author}{\bibinfo{person}{A.~Z. Broder}.}
  \bibinfo{year}{1997}\natexlab{}.
\newblock \showarticletitle{On the Resemblance and Containment of Documents}.
  In \bibinfo{booktitle}{{\em Proceedings of the Compression and Complexity of
  Sequences}}. \bibinfo{pages}{21--29}.
\newblock


\bibitem[\protect\citeauthoryear{Cormen, Leiserson, Rivest, and Stein}{Cormen
  et~al\mbox{.}}{2009}]%
        {Cormen2009}
\bibfield{author}{\bibinfo{person}{T.~H. Cormen}, \bibinfo{person}{C.~E.
  Leiserson}, \bibinfo{person}{R.~L. Rivest}, {and} \bibinfo{person}{C.
  Stein}.} \bibinfo{year}{2009}\natexlab{}.
\newblock \bibinfo{booktitle}{{\em Introduction to Algorithms}}.
\newblock \bibinfo{publisher}{MIT Press}.
\newblock


\bibitem[\protect\citeauthoryear{Dahlgaard, Knudsen, and Thorup}{Dahlgaard
  et~al\mbox{.}}{2017}]%
        {Dahlgaard2017}
\bibfield{author}{\bibinfo{person}{S. Dahlgaard}, \bibinfo{person}{M.~B.~T.
  Knudsen}, {and} \bibinfo{person}{M. Thorup}.}
  \bibinfo{year}{2017}\natexlab{}.
\newblock \showarticletitle{Fast Similarity Sketching}.
\newblock  (\bibinfo{year}{2017}).
\newblock
\showeprint{1704.04370}


\bibitem[\protect\citeauthoryear{Fisher and Yates}{Fisher and Yates}{1938}]%
        {Fisher1938}
\bibfield{author}{\bibinfo{person}{R.~A. Fisher} {and} \bibinfo{person}{F.
  Yates}.} \bibinfo{year}{1938}\natexlab{}.
\newblock \bibinfo{booktitle}{{\em Statistical Tables for Biological,
  Agricultural and Medical Research}}.
\newblock \bibinfo{publisher}{Oliver \& Boyd, London}.
\newblock


\bibitem[\protect\citeauthoryear{Li, Owen, and Zhang}{Li et~al\mbox{.}}{2012}]%
        {Li2012}
\bibfield{author}{\bibinfo{person}{P. Li}, \bibinfo{person}{A. Owen}, {and}
  \bibinfo{person}{C. Zhang}.} \bibinfo{year}{2012}\natexlab{}.
\newblock \showarticletitle{One Permutation Hashing}.
\newblock \bibinfo{journal}{{\em Advances in Neural Information Processing
  Systems\/}}  \bibinfo{volume}{25} (\bibinfo{year}{2012}),
  \bibinfo{pages}{3113--3121}.
\newblock


\bibitem[\protect\citeauthoryear{Lumbroso}{Lumbroso}{2013}]%
        {Lumbroso2013}
\bibfield{author}{\bibinfo{person}{J. Lumbroso}.}
  \bibinfo{year}{2013}\natexlab{}.
\newblock \showarticletitle{Optimal Discrete Uniform Generation from Coin
  Flips, and Applications}.
\newblock  (\bibinfo{year}{2013}).
\newblock
\showeprint{1304.1916}


\bibitem[\protect\citeauthoryear{Shrivastava}{Shrivastava}{2017}]%
        {Shrivastava2017}
\bibfield{author}{\bibinfo{person}{A. Shrivastava}.}
  \bibinfo{year}{2017}\natexlab{}.
\newblock \showarticletitle{Optimal Densification for Fast and Accurate Minwise
  Hashing}.
\newblock  (\bibinfo{year}{2017}).
\newblock
\showeprint{1703.04664}


\bibitem[\protect\citeauthoryear{Shrivastava and Li}{Shrivastava and
  Li}{2014a}]%
        {Shrivastava2014}
\bibfield{author}{\bibinfo{person}{A. Shrivastava} {and} \bibinfo{person}{P.
  Li}.} \bibinfo{year}{2014}\natexlab{a}.
\newblock \showarticletitle{Densifying One Permutation Hashing via Rotation for
  Fast Near Neighbor Search}. In \bibinfo{booktitle}{{\em Proceedings of the
  31st International Conference on Machine Learning}}.
  \bibinfo{pages}{557--565}.
\newblock


\bibitem[\protect\citeauthoryear{Shrivastava and Li}{Shrivastava and
  Li}{2014b}]%
        {Shrivastava2014a}
\bibfield{author}{\bibinfo{person}{A. Shrivastava} {and} \bibinfo{person}{P.
  Li}.} \bibinfo{year}{2014}\natexlab{b}.
\newblock \showarticletitle{Improved Densification of One Permutation Hashing}.
\newblock  (\bibinfo{year}{2014}).
\newblock
\showeprint{1406.4784}


\bibitem[\protect\citeauthoryear{Stadje}{Stadje}{1990}]%
        {Stadje1990}
\bibfield{author}{\bibinfo{person}{W. Stadje}.}
  \bibinfo{year}{1990}\natexlab{}.
\newblock \showarticletitle{The Collector's Problem with Group Drawings}.
\newblock \bibinfo{journal}{{\em Advances in Applied Probability\/}}
  \bibinfo{volume}{22}, \bibinfo{number}{4} (\bibinfo{year}{1990}),
  \bibinfo{pages}{866--882}.
\newblock


\end{thebibliography}

\end{document}